\documentclass[twocolumn]{emulateapj}

\usepackage{subfigure}

\usepackage{bm}

\usepackage{amsmath, amssymb}
\usepackage{times}

\usepackage[usenames]{xcolor}
\definecolor{mpl_red}{HTML}{D62728}

\usepackage[backref, breaklinks, plainpages=false, colorlinks=true, anchorcolor=blue!50!black, citecolor=blue!50!black, linkcolor=blue!50!black, urlcolor=mpl_red, bookmarks=false]{hyperref}
\usepackage[strict]{changepage}

\begin{document}

\renewcommand*{\backref}[1]{[#1]}

\newcommand{\Majid}{\href{https:/orcid.org/0000-0002-4694-4221}{\textcolor{blue!50!black}{Walid~A.~Majid}}}
\newcommand{\Pearlman}{\href{https:/orcid.org/0000-0002-8912-0732}{\textcolor{blue!50!black}{Aaron~B.~Pearlman}}}
\newcommand{\Nimmo}{\href{https://orcid.org/0000-0003-0510-0740}{\textcolor{blue!50!black}{Kenzie~Nimmo}}}
\newcommand{\Hessels}{\href{https://orcid.org/0000-0003-2317-1446}{\textcolor{blue!50!black}{Jason~W.~T.~Hessels}}}
\newcommand{\Prince}{\href{https:/orcid.org/0000-0002-8850-3627}{\textcolor{blue!50!black}{Thomas~A.~Prince}}}
\newcommand{\Naudet}{\href{https://orcid.org/0000-0001-6898-0533}{\textcolor{blue!50!black}{Charles~J.~Naudet}}}
\newcommand{\Kocz}{\href{https:/orcid.org/0000-0003-0249-7586}{\textcolor{blue!50!black}{Jonathon~Kocz}}}
\newcommand{\Shinji}{Shinji~Horiuchi}

\newcommand{\JPL}{Jet Propulsion Laboratory, California Institute of Technology, Pasadena, CA 91109, USA; \textcolor{blue}{walid.majid@jpl.nasa.gov}}
\newcommand{\CaltechPhysics}{Division of Physics, Mathematics, and Astronomy, California Institute of Technology, Pasadena, CA 91125, USA}
\newcommand{\UA}{Anton Pannekoek Institute for Astronomy, University of Amsterdam, Science Park 904, 1098 XH Amsterdam, The Netherlands}
\newcommand{\ASTRON}{ASTRON, Netherlands Institute for Radio Astronomy, Oude Hoogeveensedijk 4, 7991 PD Dwingeloo, The Netherlands}
\newcommand{\CDSCC}{CSIRO Astronomy and Space Science, Canberra Deep Space Communications Complex, P.O.~Box~1035, Tuggeranong, ACT~2901, Australia}
\newcommand{\NDSEG}{$^{\text{6}}$~NDSEG Research Fellow.}
\newcommand{\NSF}{$^{\text{7}}$~NSF Graduate Research Fellow.}


\journalinfo{{\sc Submitted to The Astrophysical Journal Letters}}
\submitted{Submitted to The Astrophysical Journal Letters}

\shorttitle{RADIO OBSERVATIONS OF FRB~121102 WITH THE DSN}
\shortauthors{MAJID ET AL.}

\title{A Dual-band Radio Observation of FRB~121102 with the Deep Space Network \\ and the Detection of Multiple Bursts}

\author{\Majid\altaffilmark{1,2}, \Pearlman\altaffilmark{2,6,7}, \Nimmo\altaffilmark{3,4}, \Hessels\altaffilmark{3,4}, \Prince\altaffilmark{2,1}, \Naudet\altaffilmark{1}, \Kocz\altaffilmark{2}, and~\Shinji\altaffilmark{5}}

\address{
$^{\text{1}}$~\JPL \\
$^{\text{2}}$~\CaltechPhysics \\
$^{\text{3}}$~\UA \\
$^{\text{4}}$~\ASTRON \\
$^{\text{5}}$~\CDSCC}

\thanks{\NDSEG}
\thanks{\NSF}


\begin{abstract}
\label{Section:Abstract}

The spectra of repeating fast radio bursts~(FRBs) are complex and time-variable, sometimes peaking within the observing band and showing a fractional emission bandwidth of about 10--30\%.  These spectral features may provide insight into the emission mechanism of repeating fast radio bursts, or they could possibly be explained by extrinsic propagation effects in the local environment. Broadband observations can better quantify this behavior and help to distinguish between intrinsic and extrinsic effects. We present results from a simultaneous 2.25 and 8.36\,GHz observation of the repeating FRB~121102 using the 70\,m Deep Space Network~(DSN) radio telescope, DSS-43. During the 5.7\,hr continuous observing session, we detected 6 bursts from FRB~121102, which were visible in the 2.25\,GHz frequency band. However, none of these bursts were detected in the 8.36\,GHz band, despite the larger bandwidth and greater sensitivity in the higher-frequency band. This effect is not explainable by Galactic scintillation and, along with previous multi-band experiments, clearly demonstrates that apparent burst activity depends strongly on the radio frequency band that is being observed.

~\vspace{-0.2cm}

\noindent
\textit{Unified Astronomy Thesaurus concepts:}~\href{http://astrothesaurus.org/uat/2008}{\textcolor{blue}{Radio transient sources~(2008)}}

\end{abstract}


\section{Introduction}
\label{Section:Introduction}

\setcounter{footnote}{7}

Fast radio bursts (FRBs) are bright (fluence\,$\sim$\mbox{0.1--400}\,Jy\,ms), 
short duration ($\sim$$\mu$s--ms) radio pulses with dispersion measures~(DMs) that are well in excess of the expected Galactic contribution along their line of sights (see, e.g.~\citealt{Petroff+2019, Cordes+2019} for recent reviews). The DMs, which are derived from frequency-dependent delays in the arrival times of the bursts due to the passage of the radio waves through the cold plasma between the source and the observer, are used as a proxy for the distances of these bursts. The high DM values have long suggested that the sources of FRBs are located at extragalactic distances. The localization of a subset of FRBs to host galaxies at redshifts of 0.034--0.66 has confirmed the extragalactic nature of FRBs~\citep{Chatterjee+2017, Bannister+2019, Ravi+2019, Prochaska+2019, Marcote+2020}. FRBs have peak flux densities that are similar to those of radio pulsars, and their extragalactic distances imply total burst energies that are $\sim$10$^{\text{10}}$--10$^{\text{14}}$ times those of pulsars, if similar beaming fractions are assumed. There is currently no well-established progenitor theory that can explain this phenomenon, though dozens of hypotheses have been proposed (e.g., see~\citealt{Platts+2019}\footnote{See \href{http://frbtheorycat.org}{http://frbtheorycat.org.}} for a catalog of theories).

Since the initial FRB discovery by \cite{Lorimer+2007}, over a hundred distinct sources have been reported (e.g., see~\citealt{Petroff+2016}\footnote{See \href{http://frbcat.org}{http://frbcat.org.}} for a catalog). Interestingly, a subset of these sources have shown repeat bursts, which has provided an opportunity to study this enigmatic phenomenon in more detail through post-facto localization of the sources to a host galaxy (e.g.,~\citealt{Marcote+2020}), studies of burst properties (e.g.,~\citealt{Gourdji+2019}), and multi-wavelength searches for potential counterparts (e.g.,~\citealt{Scholz+2017, Scholz+2020}). Whether or not all FRBs are capable of repeating remains an active debate, though it has been argued that the high overall event rate requires that a large fraction of the population are repeaters~\citep{Ravi2019}. FRBs are now also being localized precisely using the initial burst discovery data~\citep{Bannister+2019}. This will help greatly in determining whether FRBs that have only been detected once come from a physically distinct progenitor type.

FRB~121102 is the first known repeating FRB~\citep{Spitler+2014, Spitler+2016} and has been localized to a faint dwarf galaxy at a redshift of $z$\,$=$\,0.19~\citep{Chatterjee+2017, Marcote+2017, Tendulkar+2017}. Since the discovery of FRB~121102, hundreds of bursts have been detected by the Arecibo telescope and other instruments~(e.g.,~\citealt{Gourdji+2019}). Many of these detections were made at \text{$L$-band}~(\text{1--2}\,GHz), but FRB~121102 has also been detected at a wide range of radio frequencies using various radio telescopes (e.g., with CHIME/FRB at \text{0.4--0.8}\,GHz,~\citealt{Josephy+2019}; the Green Bank Telescope~(GBT) at \text{1.6--2.4}\,GHz,~\citealt{Scholz+2016, Scholz+2017}; the NASA Deep Space Network~(DSN) 70\,m radio telescope, DSS-43, at 2.25\,GHz,~\citealt{Pearlman+2019a}; the Very Large Array~(VLA) at \text{2.5--3.5}\,GHz,~\citealt{Chatterjee+2017, Law+2017}; the Arecibo telescope at \text{4.1--4.9}\,GHz,~\citealt{Michilli+2018a}; the Effelsberg 100\,m telescope at \text{4.6--5.1}\,GHz,~\citealt{Spitler+2018}; and the GBT at \text{4-8}\,GHz,~\citealt{Gajjar+2018, Zhang+2018}).

Since the progenitor population of FRBs is still unknown, broadband and high-frequency radio observations of FRBs are important for understanding the underlying emission mechanism(s). In particular, simultaneous measurements across wide bandwidths are more robust against temporal evolution of scintillation and scattering as the interference patterns change over time because of the relative motion between the source, the scattering screen, and the observer. 

In this Letter, we present results from a simultaneous observation of FRB~121102 at 2.25 and 8.36\,GHz with the NASA DSN 70\,m telescope, DSS-43, and expand upon the initial results reported in~\citet{Pearlman+2019a}. The observation and data analysis procedures are described in Section~\ref{Section:Observation}. In Section~\ref{Section:Results}, we provide measurements of the detected bursts, including the DM, width, flux density, and fluence of each burst. In Section~\ref{Section:Discussion}, we discuss our measurements of the burst spectra, previous multi-frequency measurements of FRB~121102, the impact of intrinsic and extrinsic effects on the burst properties, and the morphologies of the brightest bursts detected during this observation.


\section{Observation and Data Analysis}
\label{Section:Observation}

We observed FRB~121102 continuously for 5.7\,hr on 2019~Sep.~06, 17:27:54~UTC (MJD~58732.727708) using DSS-43, the NASA DSN 70\,m radio telescope located at the Canberra Deep Space Communication Complex~(CDSCC) in Tidbinbilla, Australia. This observation was carried out as part of a recently initiated monitoring program of repeating FRBs at high frequencies with the DSN's large 70\,m radio telescopes. DSS-43 is equipped with cryogenically-cooled, dual circular polarization receivers, which are capable of recording data simultaneously at $S$-band and $X$-band. The center frequencies of the recorded $S$-band and $X$-band data were 2.25 and 8.36\,GHz, respectively. The $S$-band system has a bandwidth of 115\,MHz, with an effective bandwidth of $\sim$100\,MHz after masking bad channels contaminated by radio frequency interference~(RFI). The $X$-band receivers provide 450\,MHz of bandwidth, with $\sim$430\,MHz of usable bandwidth. Data from both polarization channels were simultaneously received and recorded at each frequency band with two different recorders at the site's Signal Processing Center. The primary recorder is the \text{ultra-wideband} pulsar machine, described previously in~\citet{Majid+2017}, which provides channelized power spectral densities in filterbank format with a frequency resolution of 0.98\,MHz and a time resolution of 64.5\,$\mu$s.

Data in both circular polarizations were also recorded at $S$-band using the stations's very-long-baseline interferometry~(VLBI) baseband recorder in six non-contiguous sub-bands. Each sub-band spanned 8\,MHz in bandwidth and provided a total bandwidth of 48\,MHz. The center frequency of the data was 2.24\,GHz. A detailed analysis of the baseband data will be presented in an upcoming publication. Most of the results in this Letter are derived from data obtained using the ultra-wideband pulsar machine, with the exception of the autocorrelation analysis (see Section~\ref{Section:Results}).

The data were flux calibrated by measuring the system temperature, $T_{\text{sys}}$, at both frequency bands using a noise diode modulation scheme at the start of the observation, while the antenna was pointed at zenith. The $T_{\text{sys}}$ values were corrected for elevation effects, which are minimal for elevations greater than 20 degrees.

The data processing procedures were similar to those described in previous single pulse studies of pulsars and magnetars with the DSN~(e.g.,~\citealt{Majid+2017, Pearlman+2018, Pearlman+2019b}). In each data set, we corrected for the bandpass slope across the frequency band and masked bad channels corrupted by RFI, which were identified using the PSRCHIVE software package~\citep{Hotan+2004}. We also subtracted the moving average from each data value using 0.5\,s around each time sample in order to remove low frequency temporal variability.

Next, the cleaned data were dedispersed with trial DMs between 500 and 700\,pc\,cm$^{\text{--3}}$. A list of FRB candidates with detection signal-to-noise~(S/N) ratios above 6.0 were generated using a matched filtering algorithm, where each dedispersed time-series was convolved with boxcar functions with logarithmically spaced widths between $\sim$64.5\,$\mu$s and $\sim$19.4\,ms. We used a GPU-accelerated machine learning pipeline based on the \texttt{FETCH}\footnote{See \href{https://github.com/devanshkv/fetch}{https://github.com/devanshkv/fetch.}} (Fast Extragalactic Transient Candidate Hunter) software package to determine whether or not each of these FRB candidates were astrophysical~\citep{Agarwal+2019}. The same FRB candidates were also searched for astrophysical bursts using an automated classifier~\citep{Michilli+2018c, Michilli+2018d}, after independently filtering each candidate for RFI. Both of these classification pipelines identified the bursts presented in Section~\ref{Section:Results} as genuine FRBs.

In addition, we extracted raw voltages using 4.0\,s of data centered on the arrival times of each of the two brightest bursts for the autocorrelation analysis presented in Section~\ref{Section:Results}. The data were coherently dedispersed using a DM value of 563.6\,pc\,cm$^{\text{--3}}$, the structure-optimized DM associated with the brightest burst. We then used the coherently dedispersed baseband data to form filterbanks comprised of channelized power spectral densities with temporal and spectral resolutions of 32\,$\mu$s and 31.25\,kHz, respectively. The resulting burst spectra were used to calculate autocorrelation functions~(ACFs) for each burst.


\section{Results}
\label{Section:Results}

\begin{figure*}[b]
	\centering
	\includegraphics[trim=0cm 0cm 0cm 0cm, clip=false, scale=1.0, angle=0]{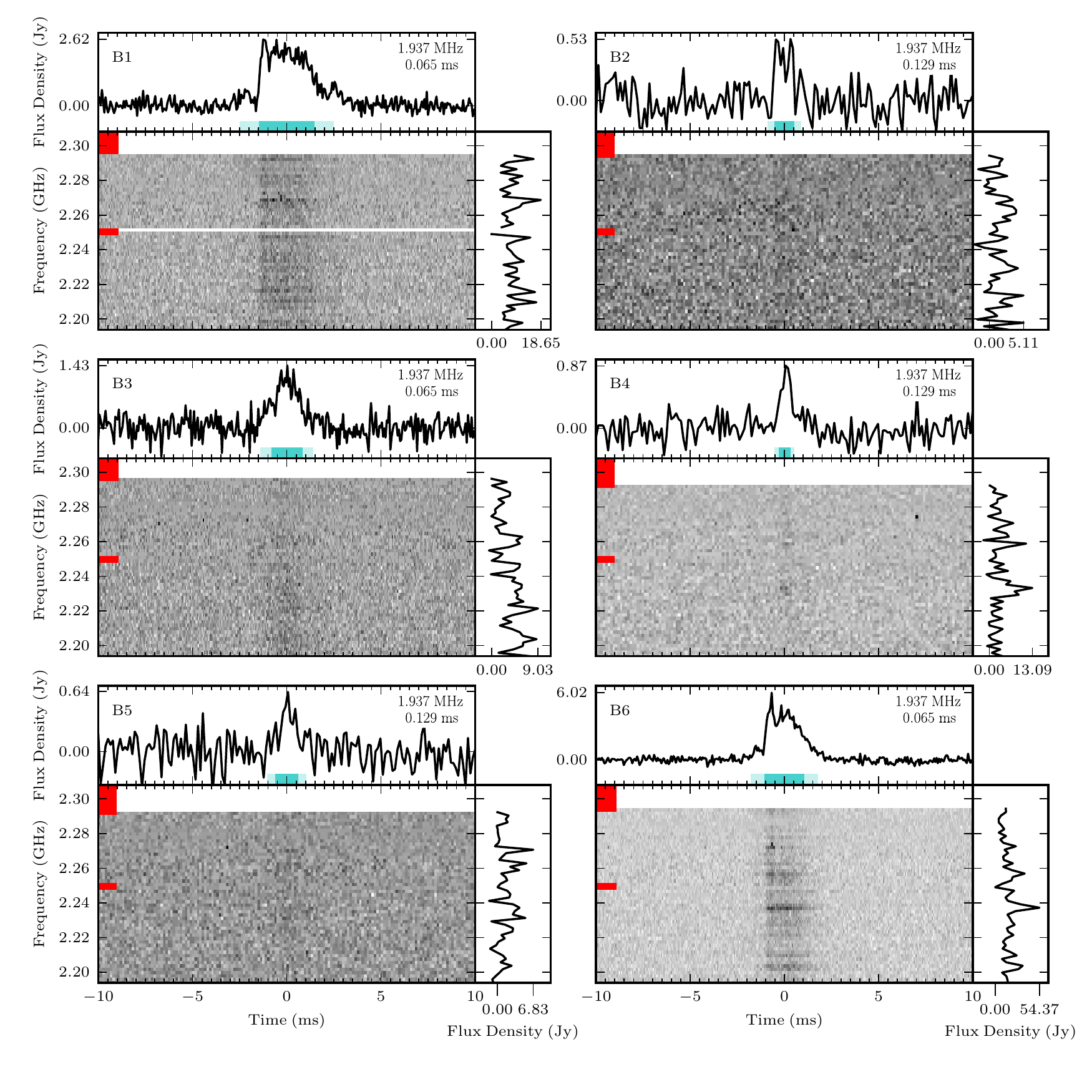}
	\caption{$S$-band bursts detected from FRB~121102 with~DSS-43, ordered by increasing arrival time. The flux calibrated, frequency-averaged burst profiles are shown in the top panels, and the dynamic spectrum associated with each burst is displayed in the bottom panels. The flux calibrated, time-averaged spectra are shown in the right panels. Each burst has been dedispersed using a DM of 563.6\,pc\,cm$^{\text{--3}}$, which corresponds to the structure-optimized DM for the brightest burst~(B6). Each burst was fitted with a Gaussian function to determine the full-width at half-maximum~(FWHM) burst duration, which is indicated with a cyan bar at the bottom of the top panels. The lighter cyan bar corresponds to a 2$\sigma$ confidence interval. The red ticks in the dynamic spectrum indicate frequency channels that have been masked as a result of~RFI. The data have been downsampled to the frequency and time resolutions specified in the top right corner of the top panels in order to enhance the visualizations of the bursts.}
	\label{Figure:Figure1}
\end{figure*}

Six bursts were detected at $S$-band with a DM value near the nominal DM of FRB~121102. In Table~\ref{Table:Table1}, we list the peak time, peak S/N, DM value that maximized the peak S/N, burst width, peak flux density, spectral energy density, and fluence for each burst. We show the flux-calibrated, frequency-averaged burst profiles, dynamic spectra, and flux-calibrated, time-averaged spectra for all of these bursts in Figure~\ref{Figure:Figure1}, after dedispersing each burst with a DM value of 563.6\,pc\,cm$^{\text{--3}}$. For the brightest bursts (B1 and B6), the structure-optimized DM value was consistent with the DM value that maximized the peak S/N. However, the algorithm\footnote{See \href{https://github.com/danielemichilli/DM_phase}{https://github.com/danielemichilli/DM\_phase.}}~\citep{Seymour+2019} used to determined the structure-optimized DM performs poorly on low S/N bursts. Therefore, we have chosen to dedisperse all of the burst spectra shown in Figure~\ref{Figure:Figure1} using the structure-optimized DM associated with the brightest burst, B6.

\begin{figure*}[b]
	\centering
	\includegraphics[trim=0cm 0cm 0cm 0cm, clip=false, scale=0.85, angle=0]{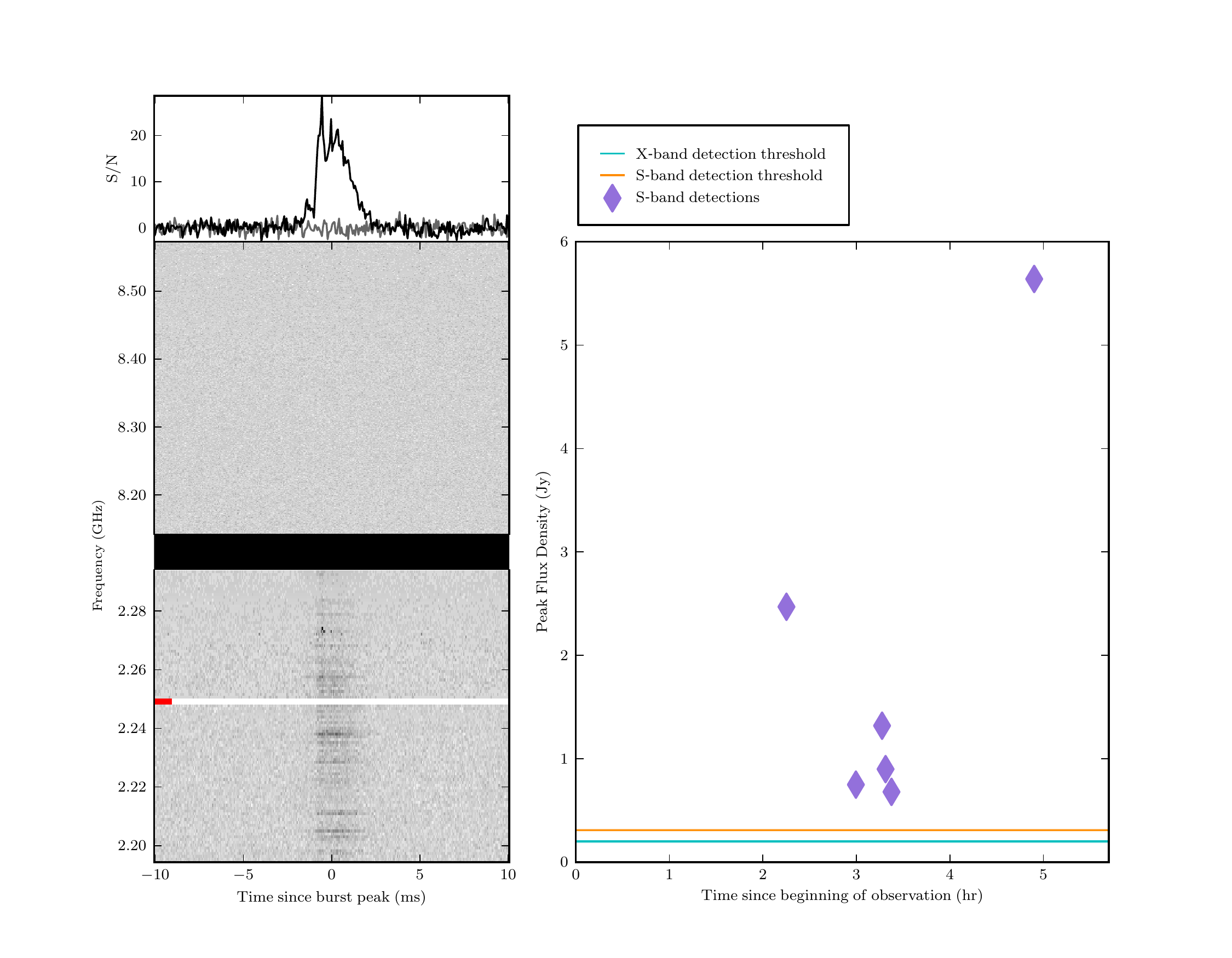}
	\caption{Left: Composite dynamic spectrum of the brightest $S$-band burst (B6), which shows the detection at $S$-band and the simultaneous non-detection at $X$-band. The time and frequency resolution plotted here are $\sim$64.5\,$\mu$s and $\sim$0.98\,MHz, respectively. The structure-optimized DM (563.6\,pc\,cm$^{\text{--3}}$) was used for dedispersion and to calculate the dispersive time delay between the $S$-band and $X$-band data. The black band indicates the frequency gap between the top of the $S$-band data and bottom of the $X$-band data. The red ticks indicate frequency channels that have been masked due to~RFI. In the top panel, we show the $S$-band frequency-averaged burst profile in black and the $X$-band frequency-averaged profile in gray. Right: Peak flux densities of the six detected $S$-band bursts as a function of time during our observation. The cyan line corresponds to the 7$\sigma$ detection threshold at $X$-band, and the orange line indicates the 7$\sigma$ detection threshold at $S$-band, both determined assuming a burst width of 1\,ms.}
	\label{Figure:Figure2}
\end{figure*}

In the left diagram in Figure~\ref{Figure:Figure2}, we show the dedispersed $S$-band and $X$-band dynamic spectra of the brightest burst, B6, after correcting for the dispersive delay between the two frequency bands. The frequency-averaged burst profiles are shown in the upper panel. Although the burst was detected with high S/N at $S$-band, there was no detectable signal during the same time at $X$-band. We also show the peak flux densities of the six detected $S$-band bursts as a function of time during our observation in the right diagram in Figure~\ref{Figure:Figure2}. The $X$-band and $S$-band 7$\sigma$ detection thresholds are indicated with cyan and orange lines, respectively. Since no bursts were detected at $X$-band, we place a 7$\sigma$ upper limit of 0.20\,Jy on the flux density of the emission at 8.36\,GHz during this observation, assuming a nominal pulse width of 1\,ms. If we further assume that the flux density scales as a power-law (i.e., $S(\nu)$\,$\propto$\,$\nu^{\alpha}$, where $S(\nu)$ denotes the flux density at an observing frequency $\nu$ and $\alpha$ is the spectral index), which is typical of most pulsar radio spectra, then we can place an upper limit of $\alpha$\,$<$\,--2.6 on the spectral index of the emission process using burst B6. However, we note that previous observations of bursts from FRB~121102 show that they may not be well-modeled by a power-law~(e.g.,~\citealt{Spitler+2016, Scholz+2016, Law+2017}).

The two brightest bursts, B1 and B6, show remarkably similar temporal profiles, with two prominent central components and a precursor component. In addition, B1 shows evidence of an additional component towards the tail of the main burst envelope. Both bursts also show spectral-temporal features that are reminiscent of other FRBs (e.g., FRB~170827;~\citealt{Farah+2018}) and other bursts from FRB~121102 (e.g.,~\citealt{Hessels+2019}).

The diffractive interstellar scintillation~(DISS) bandwidth roughly scales with frequency as:
\begin{equation}
\Delta \nu_{\text{DISS}} \propto \nu^{4}.
\label{Equation:DISS}
\end{equation}
The scintillation bandwidth of FRB~121102 was previously measured to be 58.1\,$\pm$\,2.3\,kHz at 1.65\,GHz~\citep{Hessels+2019}. The burst dynamic spectra in Figure~\ref{Figure:Figure1} show narrowband frequency structure. These structures are particularly evident in both B1 and B6 between 2.24 and 2.28 GHz. If we attribute the frequency structure in the burst spectra to DISS, then based on the scintillation bandwidth measured at 1.65\,GHz, Equation~\ref{Equation:DISS} would predict a scintillation bandwidth of $\Delta\nu_{\text{DISS}}$\,$\approx$\,200\,kHz at $\nu$\,$=$\,2.24\,GHz. We note that the data recorded from the pulsar machine is insufficient to resolve the predicted scintillation bandwidth due to its 1\,MHz spectral resolution. Therefore, to study the frequency-dependent brightness variations that arise due to scintillation, we used the baseband data to perform an ACF analysis on the burst spectra from B1 and B6. The procedure used to carry out the ACF analysis is described in detail in~\citet{Marcote+2020}. We measure the scintillation bandwidth of B1 to be 177\,$\pm$\,17\,kHz and that of B6 to be 280\,$\pm$\,13\,kHz, both at a center frequency of 2.24\,GHz. We were unable to compute ACFs for the other four bursts because they did not have sufficient~S/N.

\begin{figure*}[b]
	\centering
	\includegraphics[trim=0cm 0cm 0cm 0cm, clip=false, scale=0.95, angle=0]{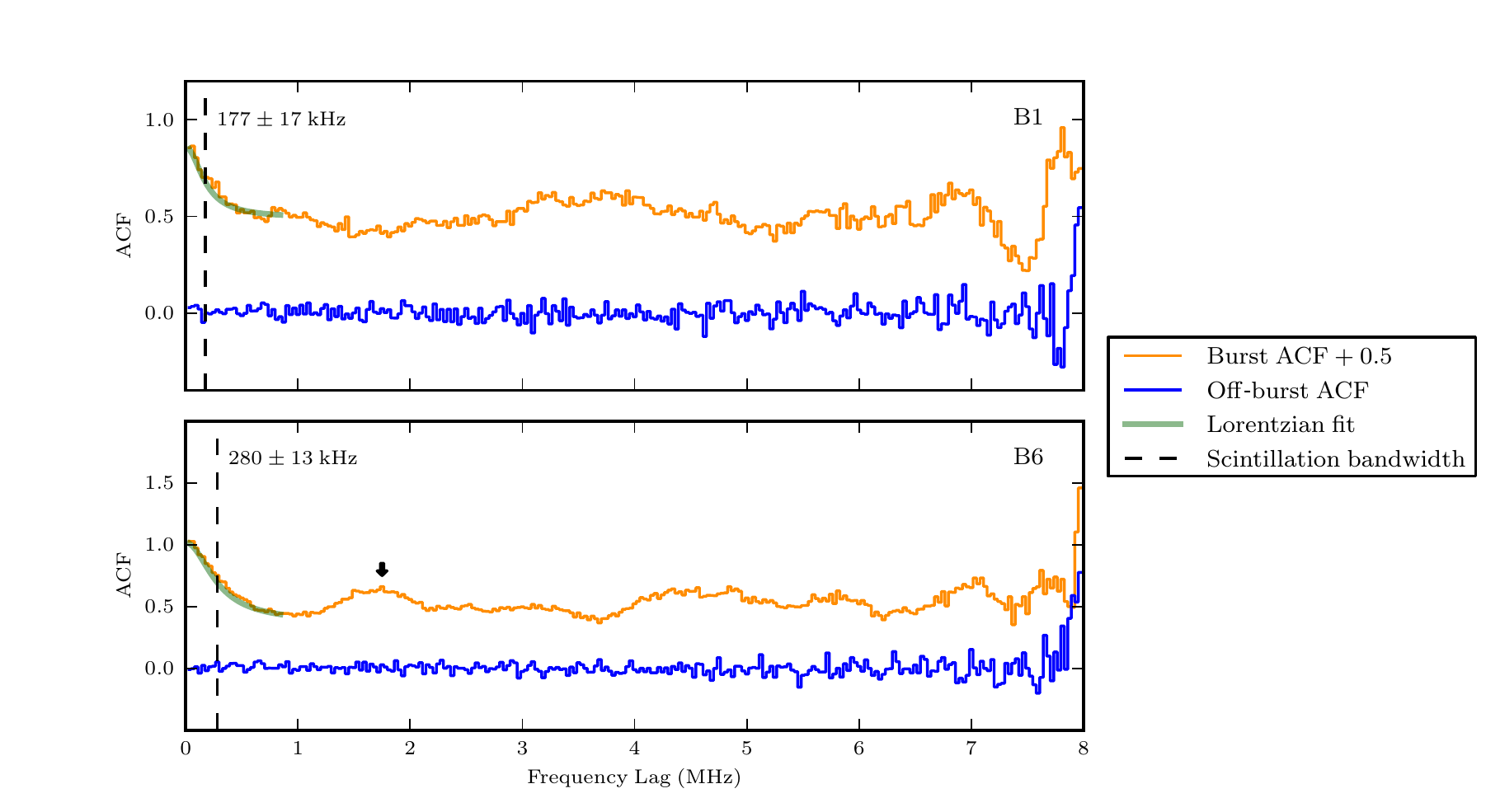}
	\caption{Autocorrelation functions~(ACFs) of the spectra associated with the two brightest $S$-band bursts, shown with frequency lags up to 8\,MHz. The ACFs are shown in orange for B1 in the top panel and for B6 in the bottom panel. The zero lag noise spike has been removed. Lorentzian fits to the central bump in the ACFs are shown in green using frequency lags up to 0.84\,MHz. The black dashed lines indicate the scintillation bandwidths, defined as the half-width at half-maximum~(HWHM) of the Lorentzian fits, and are labeled in the top left corner of each panel. The ACFs of the off-burst data are shown in blue to aid in distinguishing between frequency structure due to scintillation and instrumental effects. The black arrow in the bottom panel highlights a feature in the ACF of the spectrum of B6 at a frequency lag of $\sim$1.7\,MHz.}
	\label{Figure:Figure3}
\end{figure*}

The ACFs of both B1 and B6 are shown in Figure~\ref{Figure:Figure3} up to frequency lags of 8\,MHz. We also show Lorentzian fits to the central bump in the ACFs, which corresponds to frequency lags up to 0.84\,MHz, after removing the zero lag noise spike. The scintillation bandwidth, defined as the half-width at half-maximum~(HWHM) of the Lorentzian fit~\citep{Cordes+1985}, is labeled in the figure for each burst. The baseband data used in this analysis contained 8\,MHz frequency gaps between each of the six 8\,MHz-wide frequency sub-bands. This introduced noise spikes into the ACF at frequency lags that were close to integer multiples of the sub-band width. This is apparent in the ACFs shown in Figure~\ref{Figure:Figure3} toward frequency lags of 8\,MHz. We also include ACFs of the off-burst data in Figure~\ref{Figure:Figure3} to emphasize that the frequency structure is produced by scintillation, rather than instrumental effects. B6 shows a feature in the ACF at a frequency lag of approximately 1.7\,MHz, which we highlight using a black arrow in Figure~\ref{Figure:Figure3}. Similar behavior is not observed at the same frequency lag in the ACF of B1. We discuss this further in Section~\ref{Section:Discussion}.


\begin{deluxetable*}{ccccccc}
	\tablenum{1}
	\tabletypesize{\small}
	\tablecolumns{7}
	\tablewidth{0pt}
	\tablecaption{\textsc{Radio Bursts from FRB~121102 Detected with DSS-43}}
	\tablehead{
		\colhead{Burst ID} & 
		\colhead{Peak Time\,$^{\mathrm{a,b}}$} & 
		\colhead{DM\,$^{\mathrm{c}}$} &
		\colhead{Burst Width\,$^{\mathrm{b,d}}$} &
		\colhead{Peak Flux Density\,$^{\mathrm{b,e}}$} & 
		\colhead{Spectral Energy Density\,$^{\mathrm{b,f}}$} & 
		\colhead{Fluence\,$^{\mathrm{b,e,f}}$} \\	
		\colhead{} &
		\colhead{(MJD)} &
		\colhead{(pc\,cm$^{\text{--3}}$)} & 
		\colhead{(ms)} & 
		\colhead{(Jy)} &
		\colhead{(10$^{\text{30}}$\,erg\,Hz$^{\text{--1}}$)} &
		\colhead{(Jy\,ms)}
	}
	\startdata
	B1 & 58732.8213572248 & 564.1\,$\pm$\,0.1 & 2.94\,$\pm$\,0.06 & 2.6\,$\pm$\,0.5 & 7.5\,$\pm$\,1.5 & 6.7\,$\pm$\,1.3 \\
	B2 & 58732.8523084187 & 564.2\,$\pm$\,0.1 & 1.05\,$\pm$\,0.18 & 0.8\,$\pm$\,0.2 & 0.5\,$\pm$\,0.1 & 0.4\,$\pm$\,0.1 \\
	B3 & 58732.8639729023 & 565.0\,$\pm$\,0.1 & 1.65\,$\pm$\,0.09 & 1.4\,$\pm$\,0.3 & 2.1\,$\pm$\,0.4 & 1.8\,$\pm$\,0.4 \\
	B4 & 58732.8655320626 & 564.2\,$\pm$\,0.1 & 0.63\,$\pm$\,0.07 & 1.0\,$\pm$\,0.2 & 0.6\,$\pm$\,0.1 & 0.5\,$\pm$\,0.1 \\
	B5 & 58732.8681642140 & 564.2\,$\pm$\,0.1 & 1.23\,$\pm$\,0.19 & 0.7\,$\pm$\,0.1 & 0.6\,$\pm$\,0.1 & 0.5\,$\pm$\,0.1 \\
	B6 & 58732.9317656593 & 563.6\,$\pm$\,0.1 & 2.10\,$\pm$\,0.03 & 5.9\,$\pm$\,1.2 & 10\,$\pm$\,2.0 & 8.8\,$\pm$\,1.8
	\enddata
	\tablecomments{\\
		$^{\mathrm{a}}$ Barycentric time of the center of the burst envelope, determined after removing the time delay from dispersion using a DM value of 563.6\,pc\,cm$^{\text{--3}}$ (structure-optimized DM for the brightest burst, B6) and correcting to infinite frequency. The barycentric times were derived using the position ($\alpha_{\text{J2000}}$\,$=$\,05$^{\text{h}}$31$^{\text{m}}$58$^{\text{s}}$.698, $\delta_{\text{J2000}}$\,$=$\,33$^{\circ}$08$\arcmin$52$\arcsec$.586) in \citet{Marcote+2017}. \\
		$^{\mathrm{b}}$ Values are derived after dedispersing each burst using a DM value of 563.6\,pc\,cm$^{\text{--3}}$. \\
		$^{\mathrm{c}}$ DM value that maximized the peak S/N of each burst. \\
		$^{\mathrm{d}}$ FWHM duration determined using a Gaussian fit. \\
		$^{\mathrm{e}}$ Uncertainties are dominated by the 20\% fractional error on the system temperature, $T_{\text{sys}}$. \\
		$^{\mathrm{f}}$ Fluence determined using the 2$\sigma$ FWHM for the duration of the burst. This choice ensures that all of the burst energy is included. \\
		}
	\label{Table:Table1}
\end{deluxetable*}


\newpage

\section{Discussion and Conclusions}
\label{Section:Discussion}


To date, FRB~121102 has been detected at radio frequencies from 600\,MHz~\citep{Josephy+2019} up to 8\,GHz \citep{Gajjar+2018}. Early observations of FRB~121102 by~\citet{Spitler+2016} and~\citet{Scholz+2016} demonstrated that the bursts have variable spectra that sometimes peak within the observing band and are often not well-modeled by a power-law. This also clarifies the strange inverted spectrum of the discovery detection of FRB~121102~\citep{Spitler+2014}, though the detection of that burst in the coma lobe of the receiver likely also affected the apparent spectrum. Broader-band observations~(1.15--1.73\,GHz) by~\citet{Hessels+2019} demonstrated that the characteristic bandwidth of emission is roughly 250\,MHz at 1.4\,GHz  and the bursts are sometimes composed of sub-bursts with characteristic peak emission frequencies that decrease during the burst envelope at a rate of $\sim$200\,MHz\,ms$^{\text{--1}}$ in this frequency band. This ``sad trombone'' effect appears to be a characteristic feature of repeating FRBs~\citep{CHIME+2019}, and may be an important clue as to their emission mechanism.  The available bandwidth used to detect the 2.25\,GHz bursts presented here is insufficient to resolve sub-burst drifts of this type.

Similar narrowband, 100--200\,MHz brightness envelopes were also found by~\citet{Gourdji+2019} in a sample of 41~bursts detected using the Arecibo telescope during two $\sim$2\,hr observing sessions conducted on consecutive days. They also found tentative evidence for preferred frequencies of emission during those epochs, suggesting that FRB~121102's detectability depends strongly on the radio frequency that is being utilized. In addition, recent simultaneous, multi-frequency observations of another repeating FRB, FRB~180916.J0158+65, demonstrated that its apparent activity may also be related to the observing frequency~\citep{CHIME+2020}. CHIME/FRB detected two bursts from this source (with fluences of $\sim$2\,Jy\,ms) in the 400--800\,MHz band within a 12\,min transit. However, no bursts (above a fluence threshold of 0.17\,Jy\,ms) were detected from FRB~180916.J0158+65 with the Effelsberg telescope at $\sim$1.4\,GHz during 17.6\,hr of observations on the same day, which overlapped the times of the two CHIME/FRB detections. Clearly, the radio emission from repeating FRBs is not instantaneously broadband, which we further demonstrate with our simultaneous 2.25 and 8.36\,GHz observations of FRB~121102. Our results show that there was a period of burst activity from FRB~121102, lasting at least 2.6\,hr, where radio emission was detected at 2.25\,GHz but not at 8.36\,GHz.


There are only a few multi-band radio observations of FRB~121102 in the literature. \cite{Law+2017} present results from a multi-telescope campaign of FRB~121102 using the VLA at 3\,GHz and 6\,GHz, the Arecibo telescope at 1.4\,GHz, the Effelsberg telescope at 4.85\,GHz, the first station of the Long Wavelength Array~(LWA1) at 70\,MHz, and the Arcminute Microkelvin Imager Large Array~(AMI-LA) at 15.5\,GHz. Nine bursts were detected with the VLA, and four of these bursts had simultaneous observing coverage at different frequencies. Only one of these bursts was detected simultaneously at two different observing frequencies with Arecibo~(1.15--1.73\,GHz) and the VLA~(2.5--3.5\,GHz). The remaining three bursts were detected solely with the VLA, despite the instantaneous sensitivity of Arecibo being $\sim$5 times better than the VLA. None of the four bursts were detected during simultaneous LWA1, Effelsberg, or AMI-LA observations, though we note that only Effelsberg's sensitivity is comparable to the VLA's. \cite{Gourdji+2019} describe 41 bursts detected with Arecibo at 1.4\,GHz, and no bursts were seen with the VLA during their simultaneous observations. They also report one VLA-detected burst that was not seen in their contemporaneous Arecibo data. \cite{Houben+2019} performed a search for bursts from FRB~121102 using both Effelsberg (1.4\,GHz) and the Low Frequency Array (LOFAR; 150\,MHz). In this search, they discovered nine bursts with Effelsberg, but there were no simultaneous detections with LOFAR.


\citet{Gajjar+2018} reported the detection of 21~bursts above 5.2\,GHz during a 6\,hr observation with the~GBT. It is notable that all of these bursts were detected within a short 1\,hr time interval. The peak flux densities of these bursts ranged between $\sim$50 and $\sim$700\,mJy. These bursts also showed both large-scale ($\sim$1\,GHz-wide) and fine-scale frequency structures, none of which spanned the entire 4.5--8.0\,GHz frequency band. Assuming a flat spectral index, there are six bursts in~\citet{Gajjar+2018} with peak flux densities that are above our $X$-band sensitivity limit. Thus, similarly bright bursts would have been detected during our $X$-band observations, if they were present.


Galactic scintillation cannot explain the observed detection of bursts from FRB~121102 at $S$-band and the simultaneous absence of detection at $X$-band. However, the narrowband fluctuations of burst intensity seen at $S$-band are consistent with scintillation at low Galactic latitude ($b$\,$=$\,--0.2$^{\circ}$) expected from the Milky Way foreground ($\Delta\nu_{\text{DISS}}$\,$=$\,58.1\,$\pm$\,2.3\,kHz at 1.65\,GHz;~\citealt{Hessels+2019}). In this Letter, we have measured $\Delta\nu_{\text{DISS,\,B1}}$\,$=$\,177\,$\pm$\,17\,kHz and $\Delta\nu_{\text{DISS,\,B6}}$\,$=$\,280\,$\pm$\,13\,kHz at 2.24\,GHz for the two brightest bursts, B1 and B6.  Given the expected Galactic scintillation bandwidth of $\sim$200\,kHz and scintillation timescale of $\sim$4~minutes at 2.24\,GHz, it is not surprising that the measured scintillation bandwidths are different compared to their formal uncertainties. We are sampling a limited number of scintles in each case, and burst self-noise may also contribute to the difference. This likely also explains the other features in the on-burst ACFs, including the prominent 1.7\,MHz bump in B6.

Previously, \citet{Gajjar+2018} reported scintillation bandwidths of $\Delta\nu_{\text{DISS}}$\,$\sim$\,10--100\,MHz for bursts detected between 4.5--8.0\,GHz. Combining all available measurements, we estimate that the scintillation bandwidth is $\Delta\nu_{\text{DISS}}$\,$\approx$\,\text{0.2--0.3}\,MHz at 2.25\,GHz and $\Delta\nu_{\text{DISS}}$\,$\approx$\,\text{30--90}\,MHz at 8.36\,GHz, where the ranges correspond to assumed scalings of $\Delta\nu_{\text{DISS}}$\,$\propto$\,$\nu^{\text{4}}$ and $\Delta\nu_{\text{DISS}}$\,$\propto$\,$\nu^{\text{4.4}}$, respectively. Galactic scintillation therefore cannot explain the clear detections of the 2.25\,GHz bursts shown in Figure~\ref{Figure:Figure1} and the lack thereof in our simultaneous 8.36\,GHz data, where the bandwidth (430\,MHz) at $X$-band is many times larger than the scintillation bandwidth.


\cite{Cordes+2017} discuss the possible role of plasma lensing on the burst spectra and apparent brightness of FRBs. They argue that FRBs may be boosted in brightness on short timescales through caustics, which can produce strong magnifications ($\lesssim$\,10$^{\text{2}}$). However, we note that larger spectral gains are possible since this depends strongly on various parameters, such as the geometry of the lens, the lens' dispersion measure depth, and the scale size~\citep{Cordes+2017, Pearlman+2018}, which are currently poorly constrained. It is therefore possible that the 2.25\,GHz detections shown in Figure~\ref{Figure:Figure1} may coincide with a caustic peak.
However, it is not clear how plasma lensing could explain the downward-only frequency drifts seen in some FRBs \citep{Hessels+2019}. On the other hand, the synchrotron maser emission model from a decelerating blast wave proposed by \cite{Metzger+2019}, provides a more natural explanation for the downward-only frequency drifts in the case of a constant density model for the upstream medium. Intriguingly, this model also suggests that multiple weaker flares from the source engine in succession could produce clustered bursts over $\sim$10$^{\text{2}}$--10$^{\text{3}}$\,s by having each burst run through the same ejecta shell. We note that our sample of faint bursts, B2--B5, cluster in time $\sim$10$^{\text{3}}$\,s, while the most energetic bursts in our sample, B1 and B6, have a temporal separation of $\sim$10$^{\text{4}}$\,s.


The two brightest bursts (B1 and B6) in Figure~\ref{Figure:Figure1} display remarkably similar morphology: a weak precursor sub-burst, followed by a sharp rise and bright sub-burst (lasting for $\sim$0.5\,ms), and thereafter a broader component (lasting for a few milliseconds) perhaps composed of multiple unresolved sub-bursts, followed by a slow decay. The decaying tails of these bursts are far too long to be due to multipath propagation through the Galactic interstellar medium~(ISM), which is expected to produce a scattering time of $\tau_{d}$\,$=$\,1.16/2$\pi\Delta\nu_{\text{DISS}}$\,$\approx$\,0.7\,$\mu$s at 2.25\,GHz~\citep{Cordes+1998}. Rather, it appears that this structure may either be intrinsic to the burst emission mechanism or originate in FRB~121102's host galaxy and/or local environment. Furthermore, many other bursts from FRB~121102 also show asymmetric burst morphologies, which cannot be explained by scattering (e.g., see~\citealt{Hessels+2019}). The burst tails observed in B1 and B6 may be caused by the same mechanism responsible for the sub-burst drift rate and the apparent ``sad trombone'' behavior~\citep{Hessels+2019, Josephy+2019}.


A comparison of our structure-maximizing DM of 563.6\,$\pm$\,0.1\,pc\,cm$^{\text{--3}}$ for B6 at MJD 58732 with the reported DM\,$=$\,560\,$\pm$\,0.07\,pc\,cm$^{\text{--3}}$ at MJD 57644 by~\citet{Hessels+2019} suggests an increase of $\Delta$DM\,$\sim$\,3.6\,pc\,cm$^{\text{--3}}$ over a period of roughly 3 years. This trend agrees roughly with the $\Delta$DM\,$\sim$\,1--3\,pc\,cm$^{\text{--3}}$ in 4 years reported by~\citet{Hessels+2019}. The apparent trend, suggesting an increase in the electron column density along the line of sight, could be explained by the source moving in an H II region~\citep{Yang+2017}.  Interestingly an FRB source in an expanding SNR is expected to primarily result in a decreasing trend in DM over time, which is not borne out by the recent observations. Clearly long term observations of FRB~121102 will be needed to confirm or refute the currently observed trend.


Multi-frequency observations that densely cover the $\sim$\text{0.1--30}\,GHz range can better clarify how the burst activity of FRB~121102 depends on the radio observing frequency. It is currently unclear whether there is an optimal frequency range for observing this source. While many bursts from FRB~121102 appear to span only a few hundred MHz of bandwidth, some appear to span at least $\sim$2\,GHz~\citep{Law+2017}.  Multi-frequency, broadband measurements can also better quantify the typical emission bandwidth and determine whether or not bursts show multiple brightness peaks at widely separate frequencies, both of which are important for disentangling propagation effects and studying the mechanism(s) responsible for the emission.


\section*{Acknowledgments}

We thank Jim Cordes for useful discussions.

A.B.P. acknowledges support by the Department of Defense~(DoD) through the National Defense Science and Engineering Graduate~(NDSEG) Fellowship Program and by the National Science Foundation~(NSF) Graduate Research Fellowship under Grant~No.~\text{DGE-1144469}. J.W.T.H. acknowledges funding from an NWO Vici fellowship.

We thank the Jet Propulsion Laboratory's Spontaneous Concept Research and Technology Development program for supporting this work. We also thank Charles Lawrence and Stephen Lichten for providing programmatic support. In addition, we are grateful to the DSN scheduling team (Hernan Diaz, George Martinez, Carleen Ward) and the Canberra Deep Space Communication Complex~(CDSCC) staff for scheduling and carrying out these observations.

A portion of this research was performed at the Jet Propulsion Laboratory, California Institute of Technology and the Caltech campus, under a Research and Technology Development Grant through a contract with the National Aeronautics and Space Administration. U.S. government sponsorship is acknowledged.


\section*{ORCID iDs}
\label{Section:OrcidIDs}

\noindent
Walid~A.~Majid~\href{https:/orcid.org/0000-0002-4694-4221}{https:/orcid.org/0000-0002-4694-4221} \\
\noindent
Aaron~B.~Pearlman~\href{https:/orcid.org/0000-0002-8912-0732}{https:/orcid.org/0000-0002-8912-0732} \\
\noindent
Kenzie~Nimmo~\href{https://orcid.org/0000-0003-0510-0740}{https://orcid.org/0000-0003-0510-0740} \\
\noindent
Jason~W.~T.~Hessels~\href{https://orcid.org/0000-0003-2317-1446}{https://orcid.org/0000-0003-2317-1446} \\
\noindent
Thomas~A.~Prince~\href{https:/orcid.org/0000-0002-8850-3627}{https:/orcid.org/0000-0002-8850-3627} \\
\noindent
Jonathon~Kocz~\href{https:/orcid.org/0000-0003-0249-7586}{https:/orcid.org/0000-0003-0249-7586} \\
\noindent
Charles~J.~Naudet~\href{https://orcid.org/0000-0001-6898-0533}{https://orcid.org/0000-0001-6898-0533}



\bibliographystyle{yahapj}
\bibliography{references}


\end{document}